\documentclass[reprint,amsmath,amssymb,aps,noeprint]{revtex4-2}
\usepackage{graphicx}
\usepackage{xcolor}
\usepackage{newtxtext}
\usepackage[varvw]{newtxmath}
\usepackage[colorlinks=true,citecolor=blue,linkcolor=blue,urlcolor=blue]{hyperref}

\graphicspath{{./figs}{./}}

\begin{document}

\title{Multi-impurity method for the bond-weighted tensor renormalization group}

\author{Satoshi Morita}
\email[]{smorita@keio.jp}
\affiliation{Graduate School of Science and Technology, Keio University,
  Yokohama, Kanagawa 223-8522, Japan}
\affiliation{Keio University Sustainable Quantum Artificial Intelligence Center (KSQAIC),
  Keio University, Minato-ku, Tokyo 108-8345, Japan}

\author{Naoki Kawashima}
\affiliation{Institute for Solid State Physics, The University of Tokyo,
  Kashiwa, Chiba 277-8581, Japan}
\affiliation{Trans-scale Quantum Science Institute, The University of Tokyo,
  Bunkyo-ku, Tokyo 113-0033, Japan}


\begin{abstract}
  We propose a multi-impurity method for the bond-weighted tensor renormalization group (BWTRG) to compute the higher-order moment of physical quantities in a two-dimensional system.
  The replacement of the bond weight with an impurity matrix in a bond-weighted triad tensor network represents a physical quantity such as the magnetization and the energy.
  We demonstrate that the accuracy of the proposed method is much higher than the conventional tensor renormalization group for the Ising model and the five-state Potts model.
  Furthermore, we perform the finite-size scaling analysis and observe that the dimensionless quantity characterizing the structure of the fixed point tensor satisfies the same scaling relation in the critical region as the Binder parameter.
  The estimated critical temperature dependence on the bond dimension indicates that the exponent relating the correlation length to the bond dimension varies continuously with respect to the BWTRG hyperparameter.
  We find that BWTRG with the optimal hyperparameter is more efficient in terms of computational time than alternative approaches based on the matrix product state in estimating the critical temperature.
\end{abstract}

\maketitle

\section{Introduction}

Tensor networks effectively represent the partition function of a classical statistical system or the path integral of a lattice gauge theory.
They are also used to represent the wave function of a quantum many-body system.
The exact contraction of a tensor network is generally intractable because of the exponential growth of the computational cost.
The tensor renormalization group (TRG) method approximates the contraction by combining the real-space renormalization group concept and information compression by the truncated singular value decomposition (SVD)~\cite{levin2007tensor}.
Variants of TRG, including the higher-order TRG (HOTRG)~\cite{xie2012coarsegraining}, have been proposed,
and their accuracy and computational efficiency have been improved~\cite{xie2009second,evenbly2015tensor,yang2017loop,hauru2018renormalization,harada2018entanglement,adachi2020anisotropic,kadoh2019renormalization,morita2021global,kadoh2022triad,homma2024nuclear}.
One of recent advances is the bond-weighted tensor renormalization group (BWTRG)~\cite{adachi2022bondweighted}.
Tuning the bond weight improves the convergence to the fixed point tensor and enhances the accuracy without degrading the computational efficiency.
BWTRG has been extended to the Grassmann tensor network for fermions~\cite{akiyama2022bondweighting} and applied to the lattice gauge theory~\cite{akiyama2024tensorb}.

To understand physical properties in a system of interest, it is essential to calculate basic physical quantities such as energy and magnetization.
Representative methods for calculating physical quantities based on TRG are numerical differentiation, automatic differentiation, and the introduction of impurities.
The impurity method represents a physical quantity by changing some tensors in the tensor network~\cite{gu2008tensorentanglement}.
This method is effective when a drastic change occurs near the critical point or the singular values degenerate.
We proposed a method to calculate higher-order moments by coarse-graining a tensor network with multiple impurities~\cite{morita2019calculation}.
However, since this method is based on HOTRG, the high computational cost limits the bond dimension that could be handled.
On the other hand, the TRG-based method required special care to prevent the tensor corresponding to impurities from spreading through the tensor decomposition operation.
Thus, no TRG-based method has been proposed to calculate higher-order moments using the impurity tensor.

In this paper, we propose a method to calculate higher-order moments based on BWTRG.
Our idea is to introduce impurities by replacing the bond weight instead of the tensor.
This method has lower computational cost than the method based on HOTRG, and it is possible to perform calculations with a larger bond dimension.
It has been reported that BWTRG has higher accuracy than TRG and HOTRG even with the same bond dimension.
Therefore, it is expected that the accuracy will be improved in the analysis of phase transitions and critical phenomena using this method.

In the next section, we overview BWTRG and explain our proposed method.
To reduce the computational cost, we reformulate BWTRG with a triad tensor network.
We also give the tensor network representation of the partition function and physical quantities in the Ising model and the Potts model.
In Sec.~\ref{sec:results}, we show the results of numerical calculations and evaluate the accuracy of the proposed method.
The impurity method is suitable for finite-size scaling analysis, as it allows for the calculation of the temperature dependence of physical quantities in finite systems.
Compared to the transfer matrix method, which is widely used in tensor network approaches, finite-size scaling offers greater flexibility in analyzing higher-dimensional systems, disordered systems, and a broader range of physical quantities.
We demonstrate that combining our method with finite-size scaling analysis enables the precise determination of the critical temperature and critical exponents.
We investigate how the error in the critical temperature depends on the bond dimension and observe that the proposed method is the most computationally efficient.
The last section is devoted to discussion and conclusion.

\section{Models and Methods}
\subsection{BWTRG}

\begin{figure}[]
  \centering
  \includegraphics{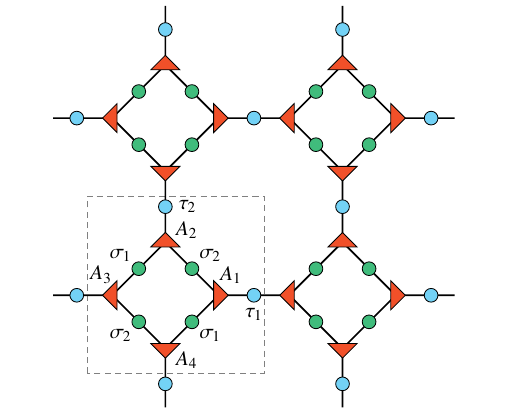}
  \caption{A bond-weighted triad tensor network on the square lattice.
  An isometric tensor and a diagonal matrix are presented by a triangle and a circle, respectively.
  The inner weight $\sigma_j$ is located on the edge of a plaquette,
  while the outer weight $\tau_j$ is on the bond connecting two plaquettes.
  A dashed square indicates a unit cell.}
  \label{fig:lattice}
\end{figure}

Let us consider a two-dimensional classical spin system under the periodic boundary condition.
The partition function at the temperature $T$ is
\begin{equation}
  Z = \sum_{\{S_x\}} e^{-H(\{S_x\})/T},
\end{equation}
where $J$ is the nearest-neighbor coupling constant, $S_x$ is a spin variable on a site $x$, and $H(\{S_x\})$ is the Hamiltonian of the system.
The summation over spin configurations can be represented as the contraction of a tensor network in various ways~\cite{zhao2010renormalization}.
However, the exact contraction is still intractable for a large system size because of the exponential growth of the computational cost.
TRG and its variants are developed to approximate the contraction by coarse-graining the tensor network.

In this paper, we consider a bond-weighted tensor network on the truncated square lattice depicted in Fig.~\ref{fig:lattice}.
We note that any two-dimensional periodic system can be transformed into a tensor network on the square lattice~\cite{zhao2010renormalization} and it can be converted to this form by one-time TRG-like SVD.
A three-index tensor on each vertex of a small square (plaquette) is denoted by $A_i$ ($i=1, \dots, 4$), which is labeled according to the orientation.
We assume that $A_i$ is isometric such as
\begin{equation}
  \includegraphics{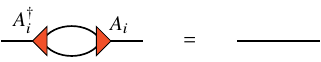},
  \label{eq:isometry_A}
\end{equation}
in the graphical representation.
Each bond has a bond weight depicted by a circle, which is assumed to be a diagonal non-negative matrix.
The bond weights can be classified into two types.
One is on a edge of a plaquette with $A_i$ at the vertices and the other is on a bond connecting those plaquettes.
We call the former an inner weight $\sigma_j$ and the latter an outer weight $\tau_j$ ($j=1, 2$).
A unit cell of this tensor network network can have four inner weights in general,
but only two independent ones remain after a renormalization step, as will become clear later.
Contracting isometric tensors and inner weights on the edges of a plaquette derives the original tensor network used in Ref.~\cite{adachi2022bondweighted}, which consists of the outer weights and the four-index tensor.
Avoiding the creation of the four-leg tensor reduces the computational cost of BWTRG as well as TRG~\cite{morita2018tensor}.

\begin{figure}[]
  \centering
  \includegraphics{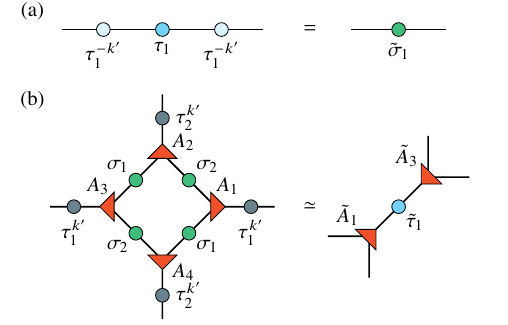}
  \caption{The update rule of BWTRG. Other components not shown here are similarly obtained by 90-degree rotated diagrams.
  (a) The renormalized inner weight is the $k$th power of the outer weight [Eq.~\eqref{eq:update_sigma}].
  (b) The truncated SVD of the left-hand side yields the outer weight and the isometric tensors in the next step.
  }
  \label{fig:update}
\end{figure}

In each step, BWTRG coarse-grains the network with a scaling factor $\sqrt{2}$ and produces a network with the same structure tilted by 45 degrees.
The inner weight in the next step is given by the $k$th power of the current outer weight.
For later convenience, we define $k' \equiv (1-k)/2$ and rewrite the update rule [Fig.~\ref{fig:update}(a)] as
\begin{equation}
  \tilde{\sigma}_j = \tau_j^k = \tau_j^{-k'} \tau_j \tau_j^{-k'}.
  \label{eq:update_sigma}
\end{equation}
Hereafter, the tilde indicates a component in the next step.
The outer weights and the isometric tensors in the next step are computed from the truncated SVD of the remaining part [Fig.~\ref{fig:update}(b)].
We keep the $\chi$ largest singular values and truncate the others.
The update rule for the other components not shown in Fig.~\ref{fig:update} is similarly obtained by 90-degree rotated diagrams.

Suppose the initial tensor network has $2^{t/2} \times 2^{t/2}$ unit cells and its contraction represents the partition function of a two-dimensional classical spin system under the periodic boundary condition.
After $t$ steps of BWTRG, the renormalized tensor network contains only one unit cell and its trace approximates the partition function as
\begin{equation}
  \includegraphics{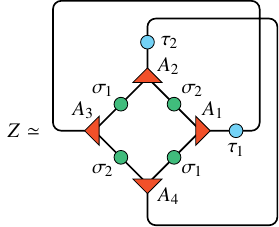}.
  \label{eq:trace}
\end{equation}
If the initial unit cell includes $N_0$ spins, the total number of spins becomes $N=2^{t} N_0$.

The hyperparameter $k$ in Eq.~\eqref{eq:update_sigma} represents the deviation from TRG and tunes the accuracy of BWTRG.
When $k=0$, the inner bond weight disappears and BWTRG is equivalent to TRG.
The authors of Ref.~\cite{adachi2022bondweighted} reported that $k=-1/2$ is optimal.
It is shown numerically that the error in the free energy is minimized at this value.
The dimensional analysis of the fixed point tensor also supports it.
In this paper, we will fix the hyperparameter $k$ to this optimal value if not mentioned.

\subsection{Impurity method}
The impurity method of tensor networks is a technique for calculating the expected value of a physical quantity,
\begin{equation}
  \langle O \rangle
  = \frac{\sum_{\{S_x\}} O(\{S_x\}) e^{-H(\{S_x\})/T}}{Z}.
\end{equation}
If the physical quantity $O$ is local, the numerator is represented by the tensor network where some of tensors are replaced with others.
The contraction of this tensor network with impurities is approximated in a similar manner to the partition function.
To calculate the spacial average of a local quantity or its higher-order moments, we proposed the multi-impurity method for HOTRG, where impurity tensors are updated by the systematic summation technique~\cite{morita2019calculation}.

Our main proposal in this paper is that we replace some of the bond weights with matrices to represent impurities.
The replaced matrix does not need to be diagonal unlike the bond weight.
There are two types of the impurity matrix as well as the bond weight, inner and outer.
We call the matrix placed at the position of the inner (outer) weight the inner (outer) impurity matrix,
which is denoted by $S_j$ ($T_j$).

\begin{figure}[]
  \centering
  \includegraphics{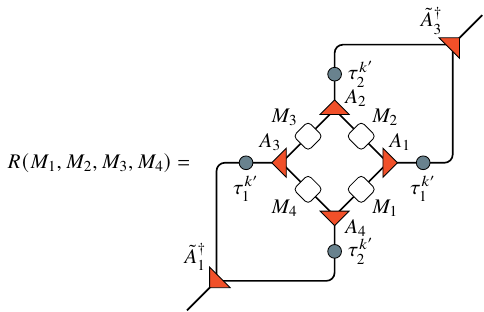}
  \caption{Diagram for calculating the outer impurity matrix $\tilde{T}_1$ in the next step.
  The matrix $M_i$ is replaced by the inner impurity matrix $S_i$ or the inner weight $\sigma_i$ at the current step.}
  \label{fig:update-imp}
\end{figure}

The update rule for the impurity matrix is similar to that for the bond weight.
The inner impurity matrix $\tilde{S}_j$ in the next step is obtained from the current outer impurity matrix $T_j$ as
\begin{equation}
  \tilde{S}_j = \tau_j^{-k'} T_j \tau_j^{-k'},
\end{equation}
which corresponds to Eq.~\eqref{eq:update_sigma}.
For the update rule for the outer impurity matrix $\tilde{T}_1$ between $\tilde{A}_1$ and $\tilde{A}_3$, we define a multilinear map $R(M_1, M_2, M_3, M_4)$, as in the diagram shown in Fig.~\ref{fig:update-imp}.
Clearly, we have $\tilde{\tau}_1=R(\sigma_1, \sigma_2, \sigma_1, \sigma_2)$ because $\tilde{\tau}_1$ satisfies the update rule shown in Fig.~\ref{fig:update}(b) and $\tilde{A}_i$ is isometric [Eq.~\eqref{eq:isometry_A}].
Depending on the position of impurities, we replace $\sigma_j$ with the inner impurity matrix $S_j$.
For example, when an impurity is on the bottom right bond, the renormalized outer impurity matrix is calculated as $R(S_1, \sigma_2, \sigma_1, \sigma_2)$.
Similarly, the other outer impurity matrix $\tilde{T}_2$ on the bond between $\tilde{A}_2$ and $\tilde{A}_4$ is updated by the 90-degree rotated diagram of Fig.~\ref{fig:update-imp}.

An impurity matrix representing the spacial average of a local physical quantity or its higher-order moment can be calculated by the systematic summation technique like as the multi-impurity method for HOTRG~\cite{morita2019calculation}.
Let $S_j[O^n]$ and $T_j[O^n]$ be the inner and outer impurity matrices for the $n$th-order moment of the physical quantity $O$.
Hereafter, for simplicity, we will omit the subscript indicating the position if not ambiguous.
The update rule of the inner impurity matrix is of the same form for any order of the moment,
\begin{equation}
  \tilde{S}[O^n] = \tau^{-k'} T[O^n] \tau^{-k'}.
\end{equation}
The update rule of the outer impurity matrix depends on the order of the moment.
We need to place the $n$ impurities at the position of $M_j$ in Fig.~\ref{fig:update-imp} and average over all possible configurations.
The outer impurity tensor for the first moment is given as the average of four possible configurations,
\begin{equation}
  \begin{split}
    \tilde{T}[O] = \frac{1}{4} & \bigl[
      R(S[O], \sigma, \sigma, \sigma)
      + R(\sigma, S[O], \sigma, \sigma) \\
      & + R(\sigma, \sigma, S[O], \sigma)
      + R(\sigma, \sigma, \sigma, S[O])
    \bigr].
  \end{split}
  \label{eq:update-imp1}
\end{equation}
For the second moment, two impurities are put on two of four possible positions or on the same position.
Thus, $\tilde{T}[O^2]$ is calculated by the weighted average of ten diagrams,
\begin{equation}
  \begin{split}
    &\tilde{T}[O^2] = \\ & \frac{1}{16} \bigl[
      2 R(S[O], S[O], \sigma, \sigma)
      + 2 R(S[O], \sigma, S[O], \sigma)\\
      & + 2 R(S[O], \sigma, \sigma, S[O])
      + 2 R(\sigma, S[O], S[O], \sigma) \\
      & + 2 R(\sigma, S[O], \sigma, S[O])
      + 2 R(\sigma, \sigma, S[O], S[O]) \\
      & + R(S[O^2], \sigma, \sigma, \sigma)
      + R(\sigma, S[O^2], \sigma, \sigma) \\
      & + R(\sigma, \sigma, S[O^2], \sigma)
      + R(\sigma, \sigma, \sigma, S[O^2])
    \bigr].
  \end{split}
  \label{eq:update-imp2}
\end{equation}
If types of impurities are more than one, we need to consider more diagrams, but its generalization is straightforward.
Our proposed method can also compute the static structure factor by carefully choosing the coefficients of each term in Eqs.~\eqref{eq:update-imp1} and \eqref{eq:update-imp2}.

\subsection{Ising model}

First, we consider the Ising model without the external magnetic field on the square lattice.
The Hamiltonian of this model is given as
\begin{equation}
  H(\{S_x \}) = -J \sum_{\langle xy \rangle} S_x S_y,
\end{equation}
where the Ising spin variable $S_x$ at a site $x$ takes $+1$ or $-1$ and the summation is taken over all neighboring site pairs.
As is well known, this model is exactly solvable at any temperature~\cite{baxter2007exactly,mccoy2014twodimensional}.
In the thermodynamic limit, the continuous phase transition occurs at the critical temperature $T_\text{c} = 2J/\ln(1+\sqrt{2})$.
The ferromagnetic phase appears below $T_\text{c}$, in which the $\mathbb{Z}_2$ spin-flip symmetry is spontaneously broken.

There are several ways to obtain a tensor network representation of the partition function.
The simplest way is to consider that information of the Ising spin is carried by the index of the bond.
In this paper, we consider that the original spin locates at the center of the plaquette in Fig.~\ref{fig:lattice}.
Since all bonds connected to a plaquette should represent only one spin at its center, the inner weight $\sigma_j$ is the $2\times 2$ identity matrix and the isometric tensor $A_j$ is $1$ if and only if its three indices are the same.
The outer weight $\tau_j$ corresponds the local Boltzmann weight $e^{JS_x S_y/T}$ because it connects the nearest-neighbor spin pair.
However, it is not diagonal in general and thus we need to diagonalize it.

The local Boltzmann weight is diagonalized by the orthogonal matrix,
\begin{equation}
  U = \frac{1}{\sqrt{2}}
  \begin{pmatrix}
    1 & 1 \\
    1 & -1
  \end{pmatrix},
\end{equation}
which gives the gauge transformation from the original spin basis to the $\mathbb{Z}_2$ symmetric one.
For later convenience, the indices of the matrix are labeled by $0$ and $1$, and then the matrix element of $U$ is given by
$U_{s,a} = (-1)^{sa}/\sqrt{2}$.
Hereafter, we use $s$ for the spin basis, and $a, b, c$ for the symmetric basis.
After applying the gauge transformation to all the bonds, we have the initial tensor network with the $\mathbb{Z}_2$ symmetry~\cite{singh2010tensor,singh2011tensor,hauru2022abeliantensors}.
Eventually, the initial bond weights are
\begin{equation}
  \sigma_j =
  \begin{pmatrix}
    1 & 0 \\
    0 & 1
  \end{pmatrix}, \quad
  \tau_j = \begin{pmatrix}
    2 \cosh K & 0 \\
    0 & 2 \sinh K
  \end{pmatrix},
\end{equation}
where we define a dimensionless quantity $K\equiv J/T$.
An element of the initial isometry after the gauge transformation of all bonds is given as
\begin{equation}
  (A_j)_{abc} = \sum_{s=0, 1} U_{s,a} U_{s,b} U_{s,c},
\end{equation}
which takes $1/\sqrt{2}$ if the $\mathbb{Z}_2$ parity is conserved; that is, is $a+b+c$ is even, and zero otherwise.
The unit cell of the initial tensor network contains one spin, that is, $N_0=1$.

We consider two physical quantities, the magnetization and the energy per site,
\begin{equation}
  m = \frac{1}{N}\sum_{x=1}^{N} S_x, \quad
  e = -\frac{J}{N}\sum_{\langle xy \rangle} S_x S_y.
\end{equation}
The initial impurity matrix for the magnetization is on the inner bond because the original spin is at the center of the plaquette in our representation, 
It is given by the gauge transformation of the Ising spin as
\begin{equation}
  S[m]
  = U^{\dagger} \begin{pmatrix}
    1 & 0 \\
    0 & -1
  \end{pmatrix} U
   = \begin{pmatrix}
    0 & 1 \\
    1 & 0
  \end{pmatrix}.
\end{equation}
One of four inner weights on the edges of the plaquette is replaced by this impurity matrix to represent the Ising spin $S_x$.
This impurity matrix has odd parity under the $\mathbb{Z}_2$ spin-flip symmetry.
Thus, the first moment of the magnetization $\left< m \right>$ is always zero and we need to calculate the second moment $\langle m^2 \rangle$.
On the other hand, the initial impurity for the energy is on the outer bond.
It includes the correlation of the nearest-neighbor spin pair in addition to the Boltzmann weight.
Thus, the gauge transformation of $-S_iS_je^{KS_i S_j}$ gives the initial outer impurity matrix for $e/J$ as
\begin{equation}
  T[e]
  = \begin{pmatrix}
    -2\sinh K & 0 \\
    0 & -2\cosh K
  \end{pmatrix}.
\end{equation}
Although it should be written precisely as $T[e/J]$, we use $J$ as the unit of energy.
This impurity matrix becomes diagonal, which reflects the fact that the matrix representation of the energy and the local Boltzmann weight have the same eigenvectors.
The initial impurity matrices of the higher-order moments are similarly given as
\begin{equation}
  \begin{gathered}
    S[m^{2n}] = \sigma, \qquad S[m^{2n+1}] = S[m], \\
    T[e^{2n}] = \tau, \qquad T[e^{2n+1}] = T[e],
  \end{gathered}
\end{equation}
because the square of the Ising spin variable is $1$, $S_x^2=1$.

\subsection{Potts model}

We also consider the ferromagnetic $q$-state Potts model~\cite{wu1982potts}.
Its Hamiltonian is given as
\begin{equation}
  H = -J \sum_{\langle xy \rangle}\delta_{S_x, S_y},
\end{equation}
where the Potts spin variable $S_x$ takes a value from $0$ to $q-1$ and $\delta_{S_x,S_y}$ denotes Kronecker's delta.
The exact transition temperature $T_c = J/\ln(1+\sqrt{q})$ is obtained by the duality relation~\cite{baxter2007exactly}.

For $q > 4$, this model shows the first-order phase transition at the transition temperature.
The exact energy just above and below $T_c$ is evaluated from the exact solution of the vertex model~\cite{baxter1973potts} as
\begin{equation}
  E(T_c\pm 0) = J\left(1 + \frac{1}{\sqrt{q}}\right)
  \left[-1 \pm \Delta(q)\tanh \left(\frac{\Theta}{2}\right)
  \right],
  \label{eq:potts_exact_energy}
\end{equation}
where $\cosh \Theta = \sqrt{q}/2$ and $\Delta(q) = \prod_{n=1}^{\infty}\tanh^2(n\Theta)$.
The latent heat is given by $L = E(T_c+0)-L(T_c-0)$.
The exact jump of the magnetization at criticality~\cite{baxter1982magnetisation} is also known as
\begin{equation}
  \Delta m = \prod_{n=1}^{\infty}\frac{1-x^{2n-1}}{1+x^{2n}},
  \label{eq:potts_exact_magnetization}
\end{equation}
where $x$ is the solution of $x + 2 + x^{-1} = q$ and $0 < x < 1$.
At $q=5$, we have $E(T_c \pm 0) / J = -1.4472 \pm 0.0265$ and $\Delta m = 0.4921$.

In the same manner as the Ising model, we construct the initial tensor network with the $\mathbb{Z}_q$ spin-rotational symmetry.
The initial inner weight $\sigma_j$ is the $q\times q$ identical matrix.
The diagonal element of the initial outer weight is the eigenvalue of the local Boltzmann weight,
\begin{equation}
  (\tau_j)_{a,a} = e^{K} - 1 + q \delta_{a,0},
\end{equation}
where the index $a$ takes a value from $0$ to $q-1$.
The local Boltzmann weight is diagonalized not by an orthogonal matrix but by a unitary matrix,
\begin{equation}
  U_{s,a} = \frac{1}{\sqrt{q}}e^{i \frac{2\pi}{q} sa}.
\end{equation}
Thus, the $\mathbb{Z}_q$ symmetric tensor network is defined on a directed graph.
Each bond has a direction which represents the flow of $\mathbb{Z}_q$ charge.
The element of the isometric tensor depends on the direction of the bond.
For example, if the indices $a$ and $b$ connect to an outgoing bond and the index $c$ connects to an incoming bond,
it is given as
\begin{equation}
  (A_j)_{abc} = \sum_{s=0}^{q-1} U_{s,a} U_{s,b} \overline{U}_{s,c}
  = \frac{1}{\sqrt{q}}\Delta_{q}(a+b-c),
\end{equation}
where the bar above $U_{s,c}$ means the complex conjugate, that is, $\bar{U}_{s, c}=U^{\dagger}_{c, s}$.
Here, $\Delta_q(a)$ is $1$ when $a$ is a multiple of $q$ and $0$ otherwise, which guarantees that $A_j$ is invariant under the global $\mathbb{Z}_q$ spin rotation.

We consider the complex magnetization as an order parameter for the Potts model to take advantage of the $\mathbb{Z}_q$ spin-rotational symmetry.
It is defined by
\begin{equation}
  m = \frac{1}{N}\sum_{x=1}^{N} e^{i\frac{2\pi}{q}S_x}.
  \label{eq:compex_magnetization}
\end{equation}
Due to the symmetry, its first moment is always zero in our calculation.
Thus, we use $\sqrt{\langle |m|^2 \rangle}$ as the order parameter of the Potts model.
Although the definition of magnetization and the boundary conditions are different from Ref.~\cite{baxter1982magnetisation} that calculated the exact value of the jump in magnetization, we can show that this magnetization \eqref{eq:compex_magnetization} also has the same jump in the thermodynamic limit.

To calculate the higher-order moment of the magnetization, $\langle |m|^n \rangle$, we need to consider two kinds of impurities, $m$ and $\overline{m}$.
Let us assume that the number of $m$ ($\overline{m}$) in a impurity matrix is $k$ ($l$).
Then, an element of the initial impurity matrix is given as
\begin{equation}
  \left(S[m^k \overline{m}^l] \right)_{a,b} = \Delta_q(-a + k - l + b),
\end{equation}
where the indices $a$ and $b$ connect to the incoming and outgoing bonds, respectively.
For example, $S[m]$ is calculated by the gauge transformation of $e^{i\frac{2\pi}{q}S_x}$.

The energy density for the Potts model is defined as
\begin{equation}
  e = -\frac{J}{N}\sum_{\langle xy \rangle} \delta_{S_x, S_y}.
\end{equation}
The initial outer impurity matrix for $e/J$ is given by the gauge transformation of
$-\delta_{S_x, S_y}e^{K\delta_{S_x, S_y}}$.
Thus, we have
\begin{equation}
  \left(T[e]\right)_{a,b} = -e^K \delta_{a, b},
\end{equation}
which is still a diagonal matrix.

\section{Numerical Results}
\label{sec:results}

\subsection{Energy and magnetization}

\begin{figure}
  \centering
  \includegraphics[width=\columnwidth]{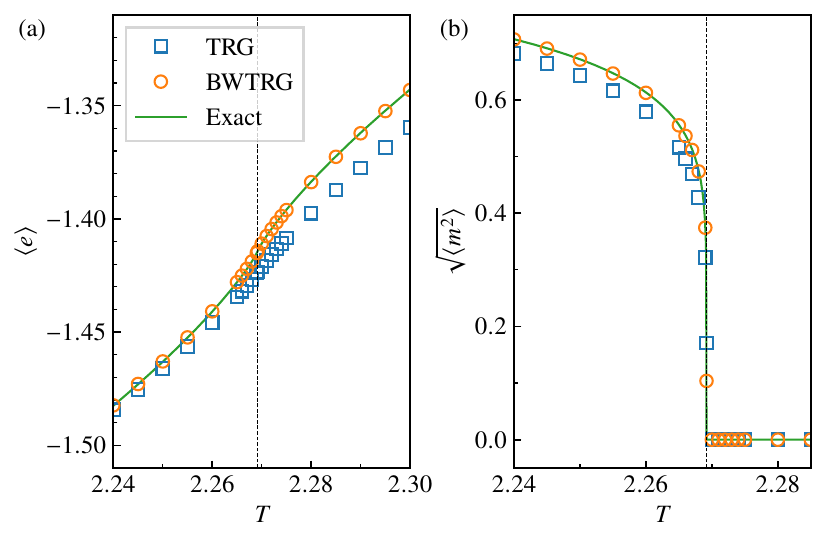}

  \caption{(a) The energy and (b) the magnetization of the Ising model on the square lattice with the system size $L=2^{25}$.
  The circle and square symbols show results by BWTRG ($k=-1/2$) and TRG ($k=0$) with the bond dimension $\chi=128$, respectively.
  The solid curves are the exact results in the thermodynamic limit and the vertical dashed lines show the exact critical temperature.
  }
  \label{fig:ising}
\end{figure}

First, we investigate the accuracy of our proposed method in the Ising model.
Figure \ref{fig:ising} shows the energy density $\langle e \rangle$ and the magnetization $\sqrt{\langle m^2 \rangle}$ calculated by our method with the bond dimension $\chi=128$.
We perform $50$ renormalization steps.
We confirm that the resulting system size $L=2^{25}$ is large enough to ignore the finite-size effect.
Although TRG ($k=0$) always underestimates both energy and magnetization, BWTRG ($k=-1/2$) is consistent with the exact result in the thermodynamic limit.

\begin{figure}
  \centering
  \includegraphics[width=\columnwidth]{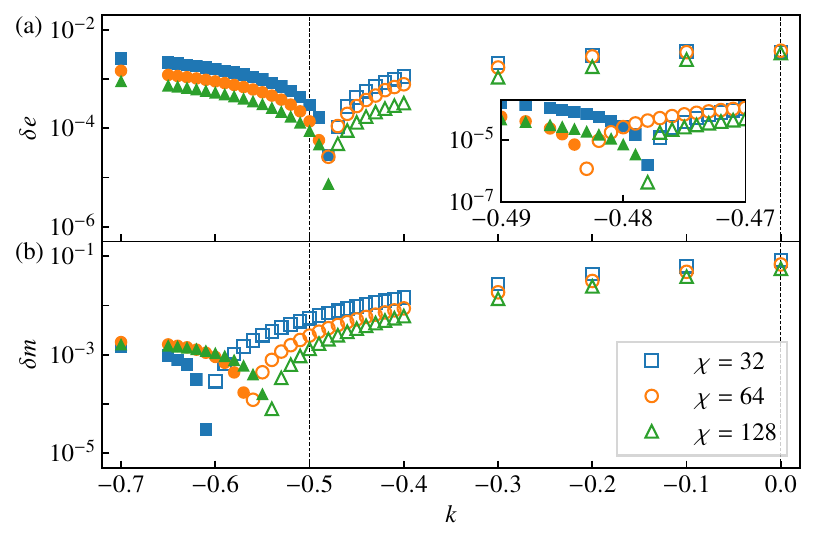}

  \caption{Relative error of (a) the energy, $\delta e = (\langle e\rangle - e_\text{exact}) / |e_\text{exact}|$,
  and (b) the magnetization, $\delta m = (\sqrt{\langle m^2\rangle} - m_\text{exact}) / m_\text{exact}$,
  of the Ising model on a square lattice with the system size $L=2^{25}$ at $T=2.26$.
  The inset shows a magnified view of the region around the energy crossing point.
  Filled (open) symbols indicate overestimation (underestimation) of a physical quantity.
  The horizontal axis is the hyperparameter $k$ in BWTRG.
  The expected optimal value of BWTRG is $k=-1/2$ and TRG corresponds to $k=0$, which are shown by the vertical dashed lines.
  }
  \label{fig:ising_vs_k}
\end{figure}

Figure~\ref{fig:ising_vs_k} shows $k$ dependence of the relative error in energy and magnetization at $T=2.26$, slightly below the critical temperature.
BWTRG with $k<0$ is always more accurate than TRG ($k=0$).
The BWTRG result crosses the exact value near the optimal value of the hyperparameter, $k=-1/2$.
The crossing point of $\langle m^2\rangle$ shifts toward $k=-1/2$ as the bond dimension increases, while that of $\langle e \rangle$ remains around $k=-0.48$.
For some $k$, the relative error exhibits nonmonotonic behavior as a function of the bond dimension because of the intersection with the exact value.
However, we find that the relative error at $k=-1/2$ decreases monotonically as the bond dimension increases.

\begin{figure}
  \centering
  \includegraphics[width=\columnwidth]{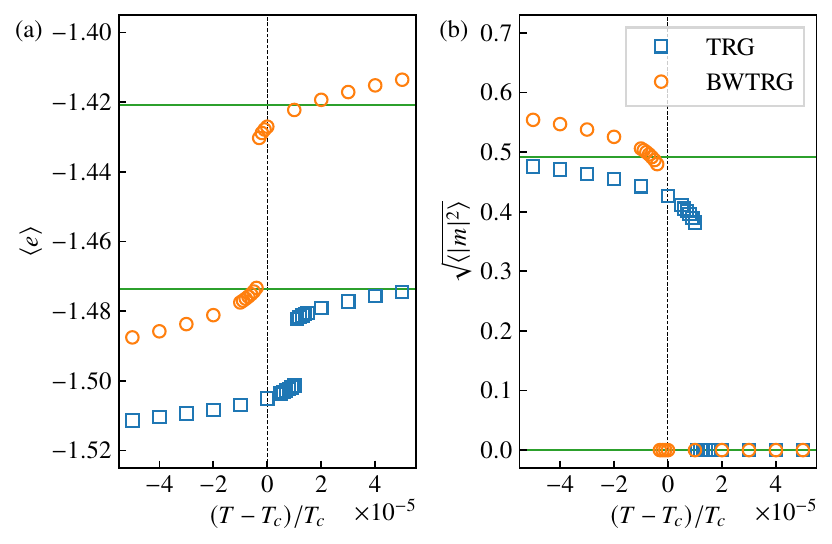}

  \caption{(a) The energy and (b) the magnetization of the five-state Potts model on the square lattice with the system size $L=2^{25}$ near the transition temperature $T_c$.
  The circle (square) symbols show results by BWTRG (TRG) with the bond dimension $\chi=150$.
  The horizontal solid lines indicate the exact values just above and below the transition temperature in the thermodynamic limit
  [Eqs.~\eqref{eq:potts_exact_energy} and \eqref{eq:potts_exact_magnetization}].
  }
  \label{fig:potts_q05}
\end{figure}

Next, we calculate the energy and the magnetization of the five-state Potts model on the square lattice as shown in Fig.~\ref{fig:potts_q05}.
Here, we perform the TRG ($k=0$) and BWTRG ($k=-1/2$) calculations with the bond dimension $\chi=150$.
Obviously, the location of the transition temperature is better estimated by BWTRG than by TRG.
The latent heat and the jump of the magnetization at the transition temperature are also improved by BWTRG.
We note that the bond dimension $\chi=150$ is much larger than that used for HOTRG in the previous work~\cite{morita2019calculation}.
It is because our BWTRG algorithm requires only $O(\chi^5)$ computational cost, while HOTRG has $O(\chi^7)$ cost.

\subsection{Finite-size scaling analysis}

The finite-size scaling analysis can estimate the critical temperature and the critical exponents.
The squared magnetization in the critical region satisfies the finite-size scaling relation,
\begin{equation}
  \left\langle m^2
  \right\rangle = L^{-2\beta/\nu} f(L^{1/\nu} t),
  \label{eq:fss_m}
\end{equation}
where $t \equiv (T-T_\text{c})/T_\text{c}$.
The exact values of critical exponents of the two-dimensional Ising universality class are $\beta=1/8$ and $\nu=1$.
The system size $L = \sqrt{N}$ is determined by the number of TRG steps as mentioned above.

A dimensionless quantity such as the Binder parameter~\cite{binder1981finite,binder1981critical} is useful in the finite-size scaling analysis because a factor before the scaling function disappears.
In this paper, we focus on the dimensionless quantity $X_1$ proposed in Ref.~\cite{gu2009tensorentanglementfiltering}.
This quantity extracts the structure of the fixed-point tensor.
In the thermodynamic limit, $X_1$ becomes $1$ in the disordered phase and $2$ in the $\mathbb{Z}_2$ symmetry breaking phase.
Thus, a jump of $X_1$ indicates the phase transition as well as the Binder parameter.
We expect its finite-size scaling relation to be in the same form as the Binder parameter,
\begin{equation}
  X_1 = g(L^{1/\nu} t).
  \label{eq:fss_X}
\end{equation}
This form follows from two facts: $X_1$ is dimensionless and the only length scales that govern $X_1$ are the correlation length and the size of the system.
Since fitting parameters in this form are only $T_\text{c}$ and $\nu$, its scaling analysis will be more stable than that of the magnetization.

The fixed-point tensor at criticality corresponds to the modular invariant partition function in the conformal field theory (CFT)~\cite{gu2009tensorentanglementfiltering}.
Thus, the universal value of $X_1$ at criticality is given as
\begin{equation}
  X_1 = \frac{\left(\sum_{\alpha} e^{-2\pi x_{\alpha}}\right)^2}
  {\sum_{\alpha} e^{-4\pi x_{\alpha}}},
  \label{eq:X1_CFT}
\end{equation}
where $\alpha$ refers to both primary operators and their descendants, and $x_\alpha$ is the scaling dimension of the scaling operator $\alpha$.
In the Ising CFT, we have $X_1 = 1.7635955$.

The definition of $X_1$ in the bond-weighted triad tensor network is straightforward.
We define a four-leg tensor $B$ by the contraction of all tensors in the unit cell (Fig.~\ref{fig:lattice}),
and $X_1$ is defined as
\begin{equation}
  X_1 = \frac{\left(\sum_{ru} B_{ruru}\right)^2}
  {\sum_{ruld} B_{rulu} B_{ldrd}},
\end{equation}
where the indices of $B$ are in counterclockwise order starting from the right leg as in Ref.~\cite{gu2009tensorentanglementfiltering}.

\begin{figure}[t]
  \centering
  \includegraphics[width=\columnwidth]{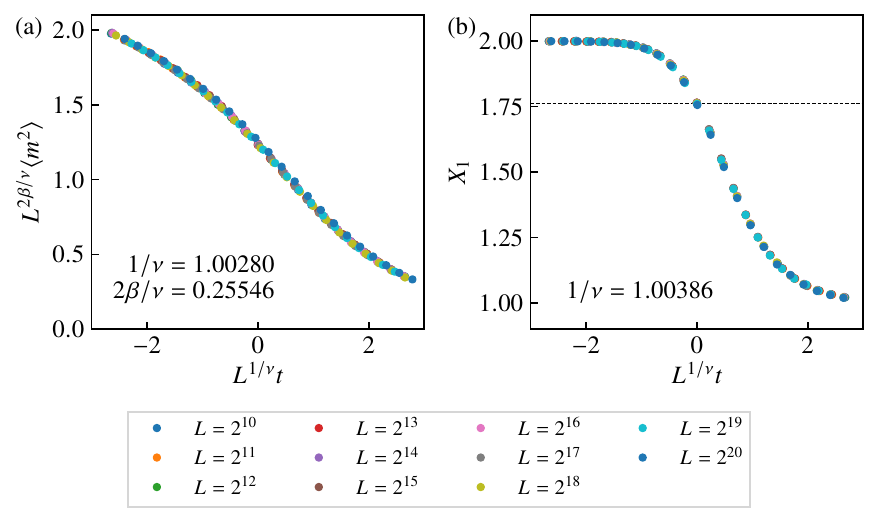}
  \caption{Finite-size scaling plots of (a) the magnetization and (b) the dimensionless quantity $X_1$ on the two-dimensional Ising model
  by BWTRG with the bond dimension $\chi=128$. The relative error in the estimated critical temperature is less than $10^{-7}$ in both plots.
  The horizontal dashed line indicates the expected value of $X_1$ at criticality from the CFT~\eqref{eq:X1_CFT}.}
  \label{fig:fss}
\end{figure}

Finite-size scaling plots of $\langle m^2 \rangle$ and $X_1$ are shown in Fig.~\ref{fig:fss}.
Here, we perform BWTRG simulations with the bond dimension $\chi=128$.
To obtain the critical exponents and the transition temperature, we use the method based on the Gaussian process regression~\cite{harada2011bayesian,harada2023fss-tools}.
All data in the wide range of the system sizes from $L=2^{10}$ to $2^{20}$ are collapsed into a scaling function.
The estimated critical exponents, $1/\nu$ and $2\beta/\nu$, agree with the exact values within relative errors of 0.39\% and 2.2\%, respectively.
The critical temperature is estimated quite accurately within a relative error of $10^{-7}$.

Figure~\ref{fig:fss}(b) clearly shows that the dimensionless quantity $X_1$ satisfies the scaling form \eqref{eq:fss_X}.
This indicates that $X_1$ takes a universal value at criticality.
Our numerical results confirm that this universal value agrees with the one predicted by the CFT analysis~\eqref{eq:X1_CFT}.
Even though BWTRG does not eliminate short-range correlations, such as the corner double-line (CDL) structure, $X_1$ successfully captures the universal properties of the fixed-point tensor.
More precisely, BWTRG delays the appearance of short-range correlation effects compared to TRG.
Once the renormalized tensor is dominated by the CDL structure, the scaling will be governed by finite entanglement rather than finite size.
From counting non-zero singular values in the full SVD of Fig.~\ref{fig:update}(b), we confirm that BWTRG with $\chi=128$ does not fall into the CDL structure even at $L=2^{25}$, while TRG with the same $\chi$ has rank reduction at $L=2^{20}$.

\begin{figure}[t]
  \centering
  \includegraphics[width=\columnwidth]{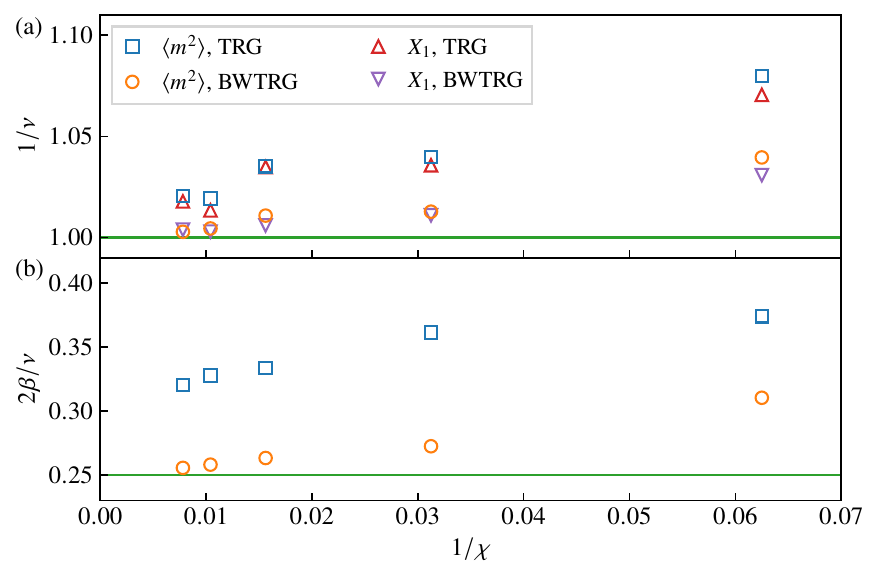}
  \caption{The $\chi$ dependence of the estimated critical exponents (a) $1/\nu$ and (b) $2\beta/\nu$ of the Ising model.
  $1/\nu$ is estimated from the magnetization \eqref{eq:fss_m} and the dimensionless quantity $X_1$ \eqref{eq:fss_X},
  while $2\beta/\nu$ is estimated from the magnetization.
  The horizontal solid lines indicate the exact values of the two-dimensional Ising universality class.}
  \label{fig:fss_exponent}
\end{figure}

The bond dimension dependence of the estimated critical exponents is shown in Fig.~\ref{fig:fss_exponent}.
As expected, the larger bond dimension gives the more accurate result.
In addition, we observe that BWTRG ($k=-1/2$) exhibits better accuracy than TRG ($k=0$).
The exponent $2\beta/\nu$ by BWTRG seems to converge to exact values in the limit of the bond dimension going to infinity, while that by TRG does not.
This is possibly due to the fact that TRG with $\chi\leq 128$ has rank reduction at $L\leq 2^{20}$.
We observe that no rank reduction occurs below $L=2^{20}$ in BWTRG with $\chi\geq 32$.

\begin{figure}[t]
  \centering
  \includegraphics[width=\columnwidth]{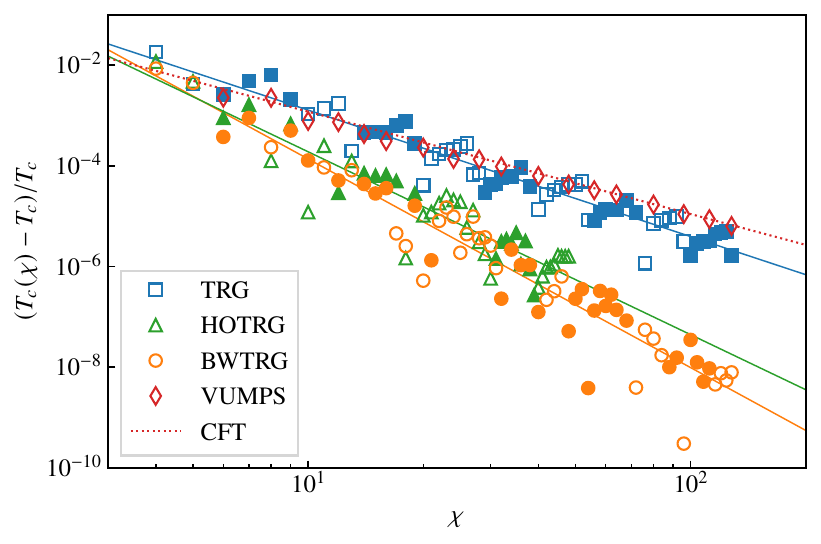}

  \caption{
    Bond-dimension dependence of the relative error in the estimated critical temperature for the Ising model on the square lattice.
    The filled and open marks indicate over- and underestimation, respectively.
    The HOTRG result is taken from Ref.~\cite{morita2019calculation}.
    The solid lines are obtained by linear fitting on the logarithmic scale for both axes.
    The VUMPS result agrees with the CFT prediction, $\delta T_\text{c} \propto \chi^{-\kappa_\text{CFT}/\nu}$, shown by the dotted line.
  }
  \label{fig:tc_err}
\end{figure}

Figure~\ref{fig:tc_err} shows the bond dimension dependence of the relative error in the estimated critical temperature $T_\text{c}(\chi)$,
\begin{equation}
  \delta T_\text{c} \equiv \frac{T_\text{c}(\chi) - T_\text{c}}{T_\text{c}}.
\end{equation}
We can robustly determine the value of $T_\text{c}(\chi)$ by finding the location of the jump in $X_1$.
We confirm that the critical temperature estimated from the jump in $X_1$ agrees with that from the finite-size scaling of the magnetization.
The estimated critical temperature $T_\text{c}(\chi)$ approaches the exact critical temperature $T_\text{c}$ with oscillation as $\chi$ increases, and the amplitude of $\delta T_\text{c}$ shows a power-law decay with respect to the bond dimension, as in HOTRG~\cite{morita2019calculation}.
We also estimate the critical temperature by using the matrix product state (MPS) approach, where the eigenvector of the transfer matrix is approximated by MPS.
We perform variational uniform MPS simulations (VUMPS)~\cite{zauner-stauber2018variational,vanderstraeten2019tangentspace}, and the critical temperature is determined from the position where the magnetization disappears.
In contrast to TRG variants, MPS results show no oscillation in the estimated critical temperature.

From the scaling theory, the effective critical temperature drifts due to the finite correlation length as $\delta T_\text{c} \sim \xi^{-1/\nu}$.
If the maximum correlation length of a tensor network with a finite bond dimension $\chi$ scales as $\xi \sim \chi^{\kappa}$, the error in the estimated critical temperature should obey the power-law behavior as
\begin{equation}
  \delta T_\text{c} \sim \chi^{-\kappa/\nu}.
  \label{eq:tc_power_law}
\end{equation}
In the MPS-base method, the CFT predicts the exponent $\kappa$ from the entanglement entropy as
\begin{equation}
  \kappa_\text{CFT} = \frac{6}{c \left(\sqrt{12/c} + 1 \right)},
  \label{eq:kappa_cft}
\end{equation}
where $c$ is the central charge~\cite{pollmann2009theory,pirvu2012matrixa}.
The Ising universality class has $c=1/2$ and it gives $\kappa_\text{CFT}=2.034$.
The dotted line in Fig.~\ref{fig:tc_err} shows that the VUMPS result is in good agreement with this CFT prediction.

\begin{figure}[t]
  \centering
  \includegraphics[width=\columnwidth]{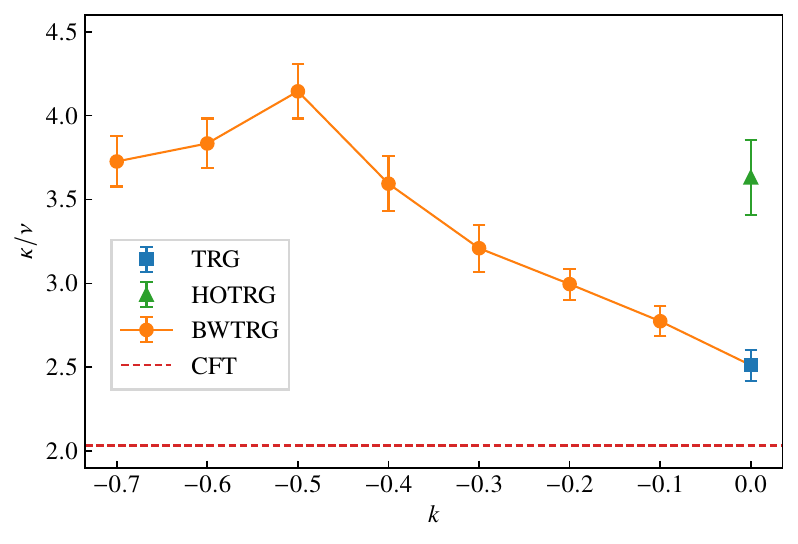}
  
  \caption{
    Dependence of the exponent $\kappa/\nu$ in Eq.~\eqref{eq:tc_power_law} on the BWTRG hyperparameter $k$.
    The horizontal dashed line is the CFT prediction obtained from $\kappa_\text{CFT}$ in Eq.~\eqref{eq:kappa_cft} and $\nu=1$.
    The error bar indicates the fitting error.
  }
  \label{fig:tc_err_exponent}
\end{figure}

Figure~\ref{fig:tc_err_exponent} shows that the exponent $\kappa/\nu$ depends on the calculation method and further on the BWTRG hyperparameter $k$.
The BWTRG method with the optimal hyperparameter $k=-1/2$ has the largest exponent, $\kappa/\nu\simeq 4.0$,
which is much larger than the CFT prediction for MPS.
Estimates from the free energy error based on $\delta f \sim \xi^{-2} \sim \chi^{-2\kappa}$ are $\kappa=2.0$ for HOTRG and $\kappa=2.2$ for BWTRG ($k=-1/2$)~\cite{ueda2014doubling,adachi2022bondweighted}.
These results are considered to provide evidence that BWTRG has the same exponent $\kappa$ as CFT~\cite{akiyama2022bondweighting}.
However, our results clearly demonstrate that the CFT prediction of $\kappa$ for MPS cannot be used in a straight-forward fashion in explaining the scaling of the correlation length in the present case.

With regard to computational time, comparison by $\kappa/\nu$ is not fair because of the different scaling of computational cost to the bond dimension.
Our BWTRG algorithm requires $O(\chi^5)$ computational cost.
Thus, error in the critical temperature is proportional to $t^{-\kappa/5\nu} \simeq t^{-0.80}$ for a calculation time $t$.
On the other hand, the minimal computational cost of MPS-based methods is proportional to $\chi^3$ and
the error in the critical temperature is proportional to $t^{-\kappa_\text{CFT}/3\nu} \simeq t^{-0.68}$.
Therefore, even taking into account the difference in computational cost, we find that BWTRG is more efficient than the MPS-based methods in estimating the critical temperature.

\section{Conclusions}\label{sec:conclusions}

In this paper, we have proposed a method to calculate physical quantities using impurities in the BWTRG method.
We have reformulated BWTRG on the truncated square lattice to reduce its computational cost to $O(\chi^5)$.
The key idea of our method is to replace the bond weight with a matrix including impurities.
The isometric tensor, which truncates unimportant information, is determined by the singular value decomposition of a part of the tensor network without impurities and made common for all physical quantities, which allows us to handle multiple impurities.
We have applied our method to the Ising model and the five-state Potts model, and demonstrated its effectiveness.
In particular, the BWTRG method gives more accurate results than the TRG method, and the error is minimized near the optimal hyperparameter $k=-1/2$.
Further improvement of BWTRG is expected by using the environment tensor~\cite{xie2009second,xie2012coarsegraining,morita2021global,kadoh2022triad} or removing the redundancy on a loop~\cite{evenbly2015tensor,yang2017loop,hauru2018renormalization,harada2018entanglement,homma2024nuclear}.
We believe that this simple impurity method is applicable to such improvement techniques.

The accuracy of the energy and magnetization (Fig.~\ref{fig:ising_vs_k}) is worse than that of the free energy, whose relative error is about $10^{-8}$ with $\chi=32$~\cite{adachi2022bondweighted}.
This is due to the fact that the proposed method does not take into account dependence of the isometric tensor on the temperature and the external magnetic field.
The automatic differentiation technique can incorporate this effect from the differentiation of the singular value decomposition, which is expected to improve the accuracy of physical quantities.
However, when the singular values are degenerate, derivatives of singular value vectors diverge, requiring special treatment.
The proposed method does not have such a problem, and it can be easily applied to any model.

We have shown that our method can be used for finite-size scaling analysis.
From the scaling analysis of the magnetization, we have estimated the critical exponents $\beta$ and $\nu$ and the transition temperature $T_\text{c}$.
We have found that the dimensionless quantity $X_1$ follows the same scaling form as the Binder parameter and have successfully estimated $\nu$ and $T_\text{c}$ (Fig.~\ref{fig:fss}).
Our findings indicate that the value of $X_1$ at criticality is universal and we have confirmed that this value agrees with the CFT arguments based on the fixed-point tensor of TRG.
The renormalized tensor network of BWTRG may include the corner double-line structure, but its redundant information is canceled out in $X_1$.
In this paper, we have performed a data-collapse analysis, but the crossing point analysis of $X_1$ is also as effective as the Binder parameter~\cite{fisher1972scaling,luck1985corrections,shao2016quantum,iino2019detecting}.

We have also investigated the bond dimension dependence of the error in the estimated critical temperature and shown that the error decreases with the power-law behavior.
Even taking into account the difference in computational cost, we have shown that the BWTRG method is more efficient than the MPS-based methods in estimating the critical temperature.
The exponent $\kappa$ of the bond dimension dependence of the effective correlation length varies continuously against the BWTRG hyperparameter $k$ (Fig.~\ref{fig:tc_err_exponent}).
This exponent is expected to reflect the entanglement structure in BWTRG as the CFT prediction for MPS, but the details are left for future research.

\begin{acknowledgments}
  We would like to thank to Synge Todo and Tsuyoshi Okubo for valuable discussions.
  The present work was supported by JSPS KAKENHI Grants No.~20K03780, No.~23H01092 and No.~23H03818.
  S.M. is supported by is the Center of Innovations for Sustainable Quantum AI (JST Grant Number JPMJPF2221).
  The computation in this work was done using the facilities of the Supercomputer Center, the Institute for Solid State Physics, the University of Tokyo.
\end{acknowledgments}

\subsection*{DATA AVAILABILITY}
The data that support the findings of this article are openly available~\cite{morita2025issp}.

\bibliography{bwtrg}
\end{document}